\documentclass[apl,aip,reprint]{revtex4-1}

\usepackage{graphicx}
\usepackage{dcolumn}
\usepackage{bm}
\usepackage{color}
\usepackage{amsfonts}

\begin{document}

\title{Evolution of locally excited avalanches in semiconductors}

\author{Z. L. Yuan}
\email{zhiliang.yuan@crl.toshiba.co.uk}

\author{J. F. Dynes}

\author{A. W. Sharpe}

\author{A. J. Shields}

\affiliation {Toshiba Research Europe Ltd, Cambridge Research Laboratory, 208 Cambridge Science Park, Milton Road, Cambridge, CB4~0GZ, UK}

\date{\today}

\begin{abstract}
We show that semiconductor avalanche photodiodes can exhibit diminutive amplification noise during the early evolution of avalanches. The noise is so low that the number of locally excited charges that seed each avalanche can be resolved. These findings constitute an important step towards realization of a solid-state noiseless amplifier for quantum information processing. Moreover, we believe that the experimental setup used, \textit{i.e.}, time-resolving locally excited avalanches, will become a useful tool for optimizing the number resolution.
\end{abstract}


\maketitle

Noiseless amplification would enable a precise single-shot readout of a quantum system through amplifying a weak signal induced by a single photon\cite{miller03} or a single charge.\cite{elzerman04} This readout ability is of prime importance in the quest of quantum information processing.\cite{knill01, kok07} Unfortunately, noise always accompanies direct amplification. For example, optical amplifiers\cite{mears87} have a minimum excess noise factor of two, ruling out their usability for discrete quanta amplification. Field-effect devices,\cite{fujiwara05,kardynal07,gansen07} based on charge-to-current conversion, have been demonstrated with single charge sensitivity, but their large timing jitter limits their prospect for practical use which requires fast single-shot measurements.

An avalanche photodiode (APD) realizes amplification through repeated impact ionization by charge carriers that have been sufficiently accelerated under high electrical field in the multiplication layer.\cite{mcintyre66} This gain ($M$) is not uniform because of the random nature of the impact ionization process.  The excess noise factor is defined as
\begin{equation}
F(M)=\frac{\langle M^2 \rangle}{\langle M \rangle^2}.
\end{equation}
\noindent  The value of $F$ quantifies the degradation in the signal-to-noise ratio after amplification. $F=1$ represents noiseless amplification, while the condition that $\frac{1}{F-1}\gg1$  must be met for discrete quanta amplification.\cite{waks03}  For most avalanche materials, the excess noise factor increases with gain and often exceeds 10 at a gain of merely $M=100$.\cite{campbell07,david08} Consequently, counting the number of individual charges through amplification using solid-state APDs has hitherto been regarded as an impossible task.

We show here, in contrast to the behavior reported for low gain, that high gain amplification during a short time window can produce nearly noiseless amplification. In our experiments, the individual carriers required to stimulate the avalanche are provided by a strongly attenuated laser pulse which is focused to a small spot on the device surface with a microscope objective. We observe that the generated signal is proportional to the number of photo-excited carriers that have stimulated the avalanche even when excitation is localized to a micron diameter spot. This demonstrates that APDs can act as photon number resolving detectors,\cite{kardynal08,wu09} without resorting to multi-pixel devices\cite{yamamoto06} or generating avalanches in different areas of a large device.\cite{waks03}

\begin{figure}
\centering\includegraphics[width=.95\columnwidth]{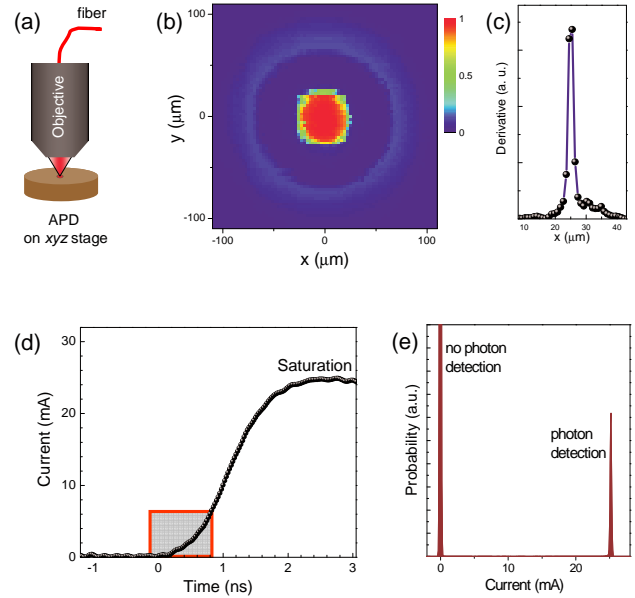}
\caption{(a) Optical setup; (b) Photocurrent as a function of $x-y$ position of the $xyz$ stage measured under low-gain operation mode; (c) The photocurrent derivative as a function of the $x$-stage position;  (d) An avalanche trace recorded by an oscilloscope; and (e) A temporal current distribution under a saturation delay.}
\label{fig:setup}
\end{figure}

Our experiment consists of a microscopic-optic setup for time-resolving locally excited avalanches.  As shown in Fig.~1(a), a microscope objective is used to focus the illumination from a 1550~nm pulsed laser into an InGaAs APD which is placed on a three-axis scanning stage. The APD has a circular active area with a diameter of 50~$\mu$m, as shown by the scanning photocurrent image (Fig.~1(b)). By scanning across the device edge (Fig.~1(c)), we determine the laser spot to have a diameter of 1.9~$\mu$m, 25 times smaller than the active diameter, thus allowing local optical excitation. Spatially localized excitation at the center of the active area is used throughout the study.

The InGaAs APD studied has a breakdown voltage of 47.6~V at a temperature of --50~$^\circ$C.  It is biased by voltage pulses of 5.6 V amplitude and 7.5 ns duration at a repetition rate of 24 kHz in combination of a DC voltage. The DC bias is set as 45.5 V, which is 2.1 V below its breakdown voltage. The 5.6~V gate pulse corresponds to an excessive bias voltage of 3.5~V.  The avalanche current is sensed by the voltage drop across a 50~$\Omega$ series resistor,\cite{yuan07} which is recorded by an oscilloscope.  The optical excitation is synchronized with the electrical signal, and their relative delay is set so that clear avalanche traces can be recorded.

Figure~1(d) shows a typical avalanche trace: the current grows gradually first, and then saturates at $25$~mA after 2~ns. The saturation inhibits all the variation in amplification gain that existed in early avalanche development. As a result, the APD behaves as a binary detector that can only resolve the absence or existence of photons.\cite{hadfield09} Little information on avalanche amplification noise can be revealed \textcolor{black}{by examining the distribution of the saturation current from the avalanches (Fig. 1(e))}.  Therefore, to study noise, we will concentrate on the initial stage of the avalanche, as labeled by the red rectangle in Fig. 1(d).

\begin{figure}
\centering\includegraphics[width=.95\columnwidth]{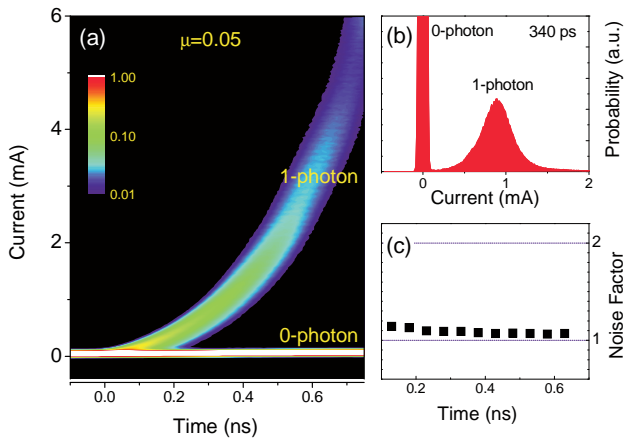}
\caption{(a) 2-D plot showing the current distribution as a function of time for a detected flux of 0.05 photons/pulse; (b) A temporal current distribution at a time delay of 340~ps;  (c) The noise factor $\mathfrak{F}(t)$ for the 1-photon avalanche as a function of time.}
\label{fig:48dB}
\end{figure}

We first study the temporal development of single photon avalanches. Of the electron-hole pair induced by a single photon absorption, only the hole can drift into the multiplication layer to initiate an avalanche due to the separation of absorption and multiplication structure used in InGaAs APDs.\cite{buller10} Therefore, these avalanches are a result of amplifying a single positive charge (hole).  To guarantee single charge amplification, the incident optical pulses are attenuated so that on average only 0.05 photons are detected per gate. Figure~2(a) shows the time development of the current distribution of single photon induced avalanches under weak illumination. Two bands are observed: (1) the ``0-photon" band around $I=0$, corresponding to gates within which no avalanche is triggered and (2) the ``1-photon" band for which an avalanche is triggered by an absorption of a single photon.  The 1-photon band emerges at zero time delay, \textit{i.e.}, soon after the illumination pulse impinges on the device. It develops into a distinct band after $t=100$ ps. Figure~2(b) shows the current distribution at a fixed delay of $340$~ps, where the 1-photon signal peaks at 0.9~mA with a distribution width of 0.42~mA. The 0-photon signal has a half width of 0.05~mA, illustrating the low electrical noise of the measurement system.

\textcolor{black}{Since an avalanche can be quenched with high time precision for a non-saturated gain,\cite{yuan07,yuan10} it is meaningful and practically useful} to define a time-dependent noise factor $\mathfrak{F}(t)$, to characterize the avalanche current distribution, as
\begin{equation}
\mathfrak{F}(t)=\frac{\langle I^2(t) \rangle}{\langle I(t) \rangle^2},
\end{equation}
\noindent where $\langle I(t) \rangle$ represents the statistical average of the avalanche current $I$ at $t$ time delay.  We point out that only the ``1-photon" band is taken into account here and thus $\mathfrak{F}$ is independent of the avalanche triggering probability. Figure~2(c) plots $\mathfrak{F}$ as a function of the avalanche development time.  $\mathfrak{F}=1.14$ is measured for very early times, and then gradually decreases to 1.07 after 200~ps avalanche development. The decrease is attributed to the diminishing influence of the electrical noise. After 200-ps, $\mathfrak{F}$ is constant at 1.07 as the avalanche signal grows further from 0.19 to 4.36~mA.

The low noise factor ($\mathfrak{F}=1.07$) measured could have one of two origins. It could indicate nearly noiseless avalanche multiplication or alternatively it could result from a process which consumes multiplication noise, such as local saturation of the avalanche current.\cite{spinelli97} Conventional wisdom favors the second explanation, but we demonstrate local saturation cannot explain our findings. Noiseless amplification allows discrete amplification, or number resolution, of locally excited charges, while local saturation\cite{waks03} can only resolve the number of spatially separated avalanches.

\begin{figure}
\centering\includegraphics[width=.8\columnwidth]{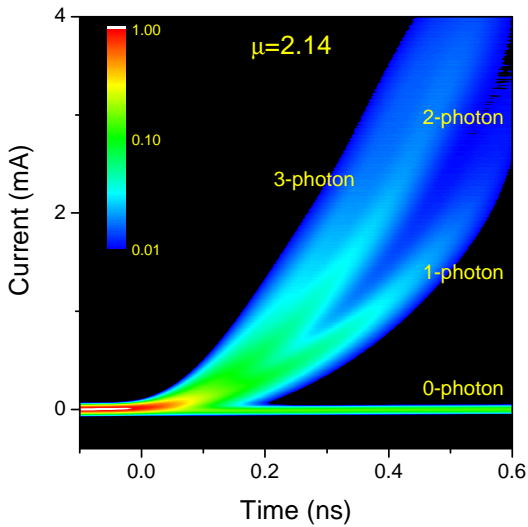}
\caption{2-D plot showing the current distribution as a function of time for a detected flux of 2.14 photons/pulse.}
\label{fig:32dB}
\end{figure}

To illustrate this we increase illumination intensity to produce two photons within the focused light spot.  Figure~3 shows the development of avalanches under this stronger, and focused, illumination.  In addition to the 0- and 1-photon bands, a new band, whose peak current is approximately twice the 1-photon band, is clearly observed. We attribute this to avalanches that were stimulated by two photo-excited holes. Similar to the 1-photon band, the 2-photon signal grows with time but at an elevated rate. A signal due to 3-photon avalanches is also observed.

The distribution evolution in the avalanche currents can be seen more clearly in Fig.~4. Before optical illumination (-10~ps), there is only one sharp peak at $I=0$, corresponding  to the 0-photon signal representing electrical noise. 40~ps after illumination, the 0-photon peak decreases considerably in height but is broadened, suggesting that avalanches are taking place, although they are still too weak to be resolved from the electrical noise. A further delay helps avalanches develop into a separate but asymmetric feature at 140~ps. They then grow to form two distinctive peaks at 190~ps.  The two peaks are due to avalanches stimulated by 1- and 2-photons, respectively.  A shoulder, on the higher current side of the distribution,  is visible and attributed to avalanches by 3 or more photons.  At 190~ps, the photon signals are well separated from the electronic noise. At further delays, the photon avalanche peaks shift to higher current, and increase in separation due to avalanche multiplication, while the 0-photon distribution remains unchanged in height.  This illustrates, without doubt, that number resolution is also observed for localized optical excitation.

\begin{figure}
\centering\includegraphics[width=.8\columnwidth]{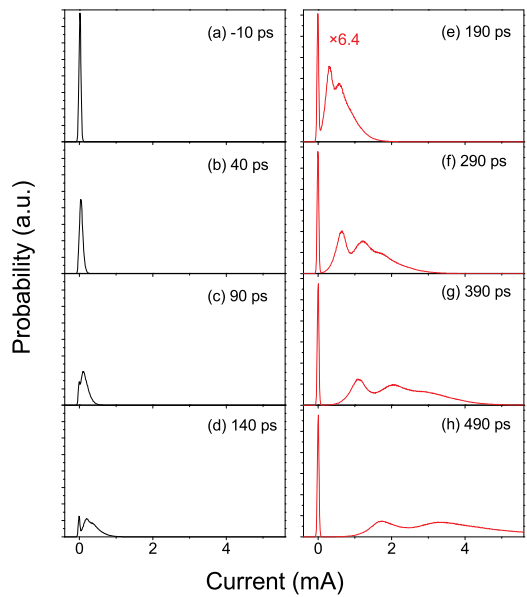}
\caption{Current distributions at various time delays for a detected flux of 2.14 photons/pulse. The vertical axes for (e--h) are scaled up by a factor of 6.4 for clarity.}
\label{fig:32dB-fig}
\end{figure}

Avalanche development involves lateral propagation. Starting from a micron-scale current filament, an avalanche is expected to propagate laterally by multiplication assisted diffusion\cite{spinelli97} until filling the entire active area of the device and reaching a saturated current. Based on the avalanche development trace in Fig.~1(d), where the current reaches to 80\% of the saturation current at a time delay of 1.5~ns, the lateral propagation is estimated in the range of 20--30~$\mu$m/ns assuming a constant propagation velocity. With such velocity, an avalanche grows into an area greater than the laser spot (1.9~$\mu$m) within 100~ps after excitation. At a greater time delay, any locally excited avalanches must have merged together, forming a single current filament.

Merging of avalanches does not affect the number resolution throughout 100 to 600~ps as shown in Fig.~3. This provides valuable insight into how an avalanche actually develops. Previously, it was suggested that an avalanche under excessive bias would undergo first a local current saturation through rapid multiplication before its lateral propagation.\cite{spinelli97} Local saturation would imply number resolution \textit{only} for spatially separated avalanches.\cite{waks03} Under a best case scenario, two separate avalanches are excited at the opposite edge of the laser spot, \textit{i.e.}, 1.9~$\mu$m apart from each other. If local current saturation did exist, a gradual loss of number resolution would be observed due to merging of individual avalanches. Clearly, this is not the case. As shown in Figs.~4(d)-(h), the number resolution does not deteriorate with time delay. In fact, the number resolution is even more clearly visible after 290~ps. The experimental results thus rule out the local saturation within the optical excitation area.  Current density must increase concurrently with lateral propagation throughout the early avalanche development.

By excluding local saturation, the noise factor ($\mathfrak{F}=1.07$) measured for the device (Fig.~2(c)) represents the upper bound of the avalanche amplification process under excessive bias.  This nearly noiseless amplification is directly responsible for the photon number resolution capabilities of the device (Figs.~3 and 4).

It is interesting to postulate about the virtual absence of avalanche multiplication noise.  It suggests a very efficient impact ionization process in the high field multiplication layer. Supposing that nearly all the energy gained by each carrier in the electrical field is used to promote further electron-hole pairs, there would be little statistical fluctuation in the avalanche multiplication. We hope our findings will stimulate further experimental and theoretical research towards a complete understanding of the multiplication noise characteristics of excessively biased APDs.

To conclude, avalanche multiplication under excessive bias occurs with little noise.  This finding opens the door towards realization of a deterministic photon or charge number detector.

%


\end{document}